\documentclass[a4paper,english]{IEEEtran}
\usepackage{graphicx}

\makeatletter

\pdfpageheight\paperheight
\pdfpagewidth\paperwidth

\providecommand{\tabularnewline}{\\}

\usepackage{subfigure}
\usepackage{amsmath}
\usepackage{mathrsfs}
\usepackage{pifont}
\usepackage{bbding}
\usepackage{tipa}
\usepackage{amsfonts}
\usepackage{amsthm}
\usepackage{amsmath,epsfig}
\usepackage{algorithm}
\usepackage{algpseudocode}
\usepackage{multicol}
\usepackage{epic}
\usepackage{url}
\usepackage{float}
\usepackage{times}
\usepackage{array}
\usepackage{mdwmath}
\usepackage{mdwtab}
\usepackage{subfigure}
\usepackage{cite}
\usepackage{amsmath}
\usepackage{amssymb}
\usepackage{footnote}
\usepackage{cmsd}
\usepackage{nccfoots}
\usepackage{curves}
\usepackage{balance}
\usepackage{epstopdf}
\usepackage{bbm}
\usepackage{color}
\usepackage{gensymb}
\usepackage{subfig}
\usepackage{tabularx}
\usepackage{multirow}
\usepackage{makecell}
\usepackage{flushend}
\makeatother

\usepackage{babel}

\usepackage{booktabs}
\usepackage{pdfpages}
\usepackage{subfigure}
\usepackage{tabularx}

\begin{document}
\title{Non-Intrusive Electric Load Monitoring Approach Based on Current Feature Visualization for Smart Energy Management}

 \author{%
	\begin{tabular}{c}
	Yiwen Xu, \emph{Member, IEEE}, Dengfeng Liu, Liangtao Huang, Zhiquan Lin, Tiesong Zhao, \emph{Senior Member, IEEE},\\ and Sam Kwong, \emph{Fellow, IEEE}\tabularnewline
	\end{tabular}

	\thanks{This work is supported by the National Natural Science Foundation of China (No. 62171134) and Foundation for Middle-aged and Young Educational Committee of Fujian Province (No. JAT200024).
		}
	\thanks{Y. Xu is with Fujian Key Lab for Intelligent Processing and Wireless Transmission of Media Information, Fuzhou University, China and also with Zhicheng College, Fuzhou University, China (e-mail:  xu\_yiwen@fzu.edu.cn).}
	\thanks{D. Liu, L. Huang and Z. Lin are with Fujian Key Lab for Intelligent Processing and Wireless Transmission of Media Information, Fuzhou University, China (e-mail: \{221120036, 21112010,  N191127023\}@fzu.edu.cn.)}
	\thanks{T. Zhao is with Fujian Key Lab for Intelligent Processing and Wireless Transmission of Media Information, Fuzhou University, China, and also with Peng Cheng Laboratory, China (e-mail: t.zhao@fzu.edu.cn).}
	\thanks{S. Kwong is with the Department of Computer Science, City University of
		Hong Kong, Kowloon, Hong Kong SAR (e-mail: cssamk@cityu.edu.hk).}

}
\maketitle

\begin{abstract}
The state-of-the-art smart city has been calling for an economic but efficient energy management over large-scale network, especially for the electric power system. It is a critical issue to monitor, analyze and control electric loads of all users in system. In this paper, we employ the popular computer vision techniques of AI to design a non-invasive load monitoring method for smart electric energy management. First of all, we utilize both signal transforms (including wavelet transform and discrete Fourier transform) and Gramian Angular Field (GAF) methods to map one-dimensional current signals onto two-dimensional color feature images. Second, we propose to recognize all electric loads from color feature images using a U-shape deep neural network with multi-scale feature extraction and attention mechanism. Third, we design our method as a cloud-based, non-invasive monitoring of all users, thereby saving energy cost during electric power system control. Experimental results on both public and our private datasets have demonstrated our method achieves superior performances than its peers, and thus supports efficient energy management over large-scale Internet of Things (IoT).
\end{abstract}


\section{Introduction}
\IEEEPARstart{T}{he} past 
decades have witnessed a booming of urban populations with ever-increased municipal facilities to serve all citizens. An effective solution to manage these facilities is smart city with Internet of Things (IoT), which is mostly benefitted from the recent development of Artificial Intelligence (AI) \cite{1,2,3}. To support the smart city, an economic but efficient electric power management system is indispensable \cite{4}. A cloud-end administrator monitors electricity consumptions of all users and loads, presents analyses of all electricity usages, and provides advices to users or directly manages electricity usage of all loads. As a result, the overall electricity consumptions are saved to support sustainable developments of cities and environments.

\begin{figure}[htb]
	\centering
	\subfigure[Invasive Load Monitoring.]{
		\includegraphics[width=3.5in]{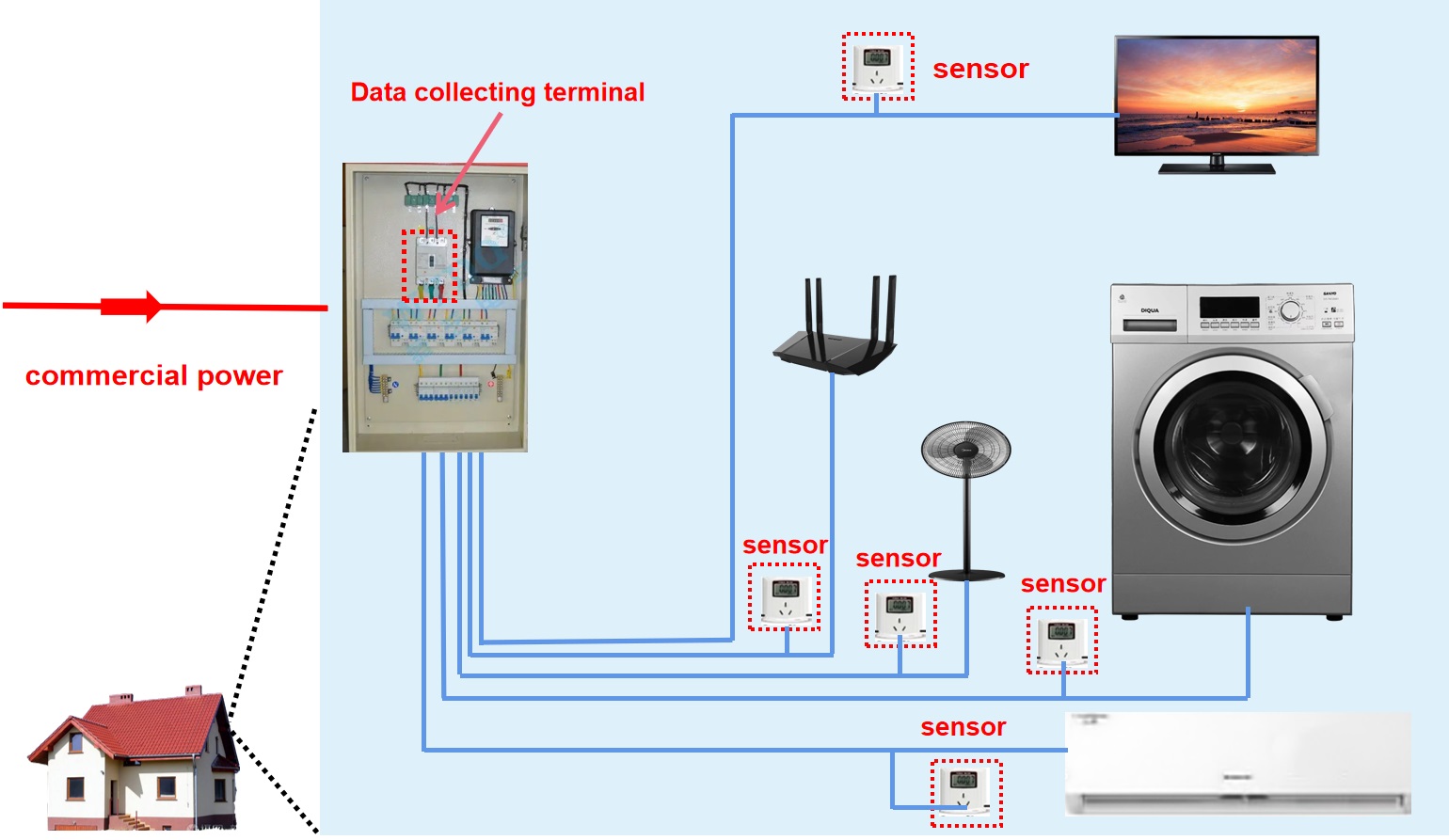}
		
	}
	
	\subfigure[Non-Invasive Load Monitoring.]{
		\includegraphics[width=3.5in]{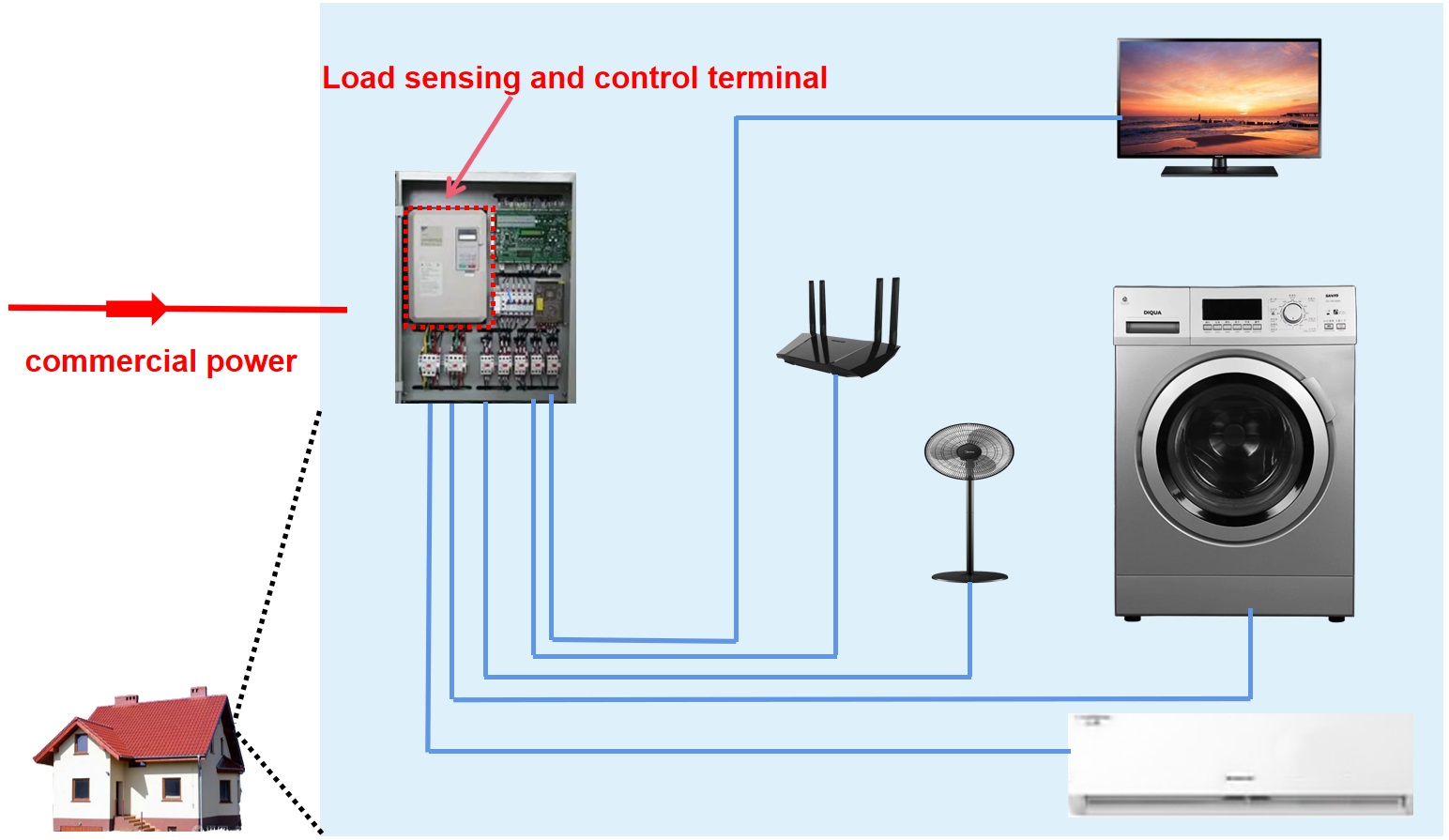}
		
	}
	
	\label{fig_1}
	\caption{The differences between invasive and non-invasive load monitoring in the electricity grid.}
\end{figure}

An efficient 
electric power management system is dependent on its electric load monitoring module \cite{5, 6, 7}, which can be realized by invasive or non-invasive approaches.  In Invasive Load Monitoring (ILM), each electric load is monitored by a separate sensor and the information acquired from all sensors can be centrally processed by cloud-end. While in Non-Invasive Load Monitoring (NILM) \cite{6, 7, 8}, only one monitor is required for each family or cell. It captures electric signals (such as voltage, current, and so on) at the commercial power input and transimits them to cloud server in which workload information of all loads are interpreted with algorithms. These differences can be shown in Fig.~1. Apparently, the non-invasive method is preferrable in smart city infrastructure for its simple design, energy-efficient and low setup/maintenance cost.

\begin{figure}[htb]
	\centering
	\includegraphics[width=3.5in]{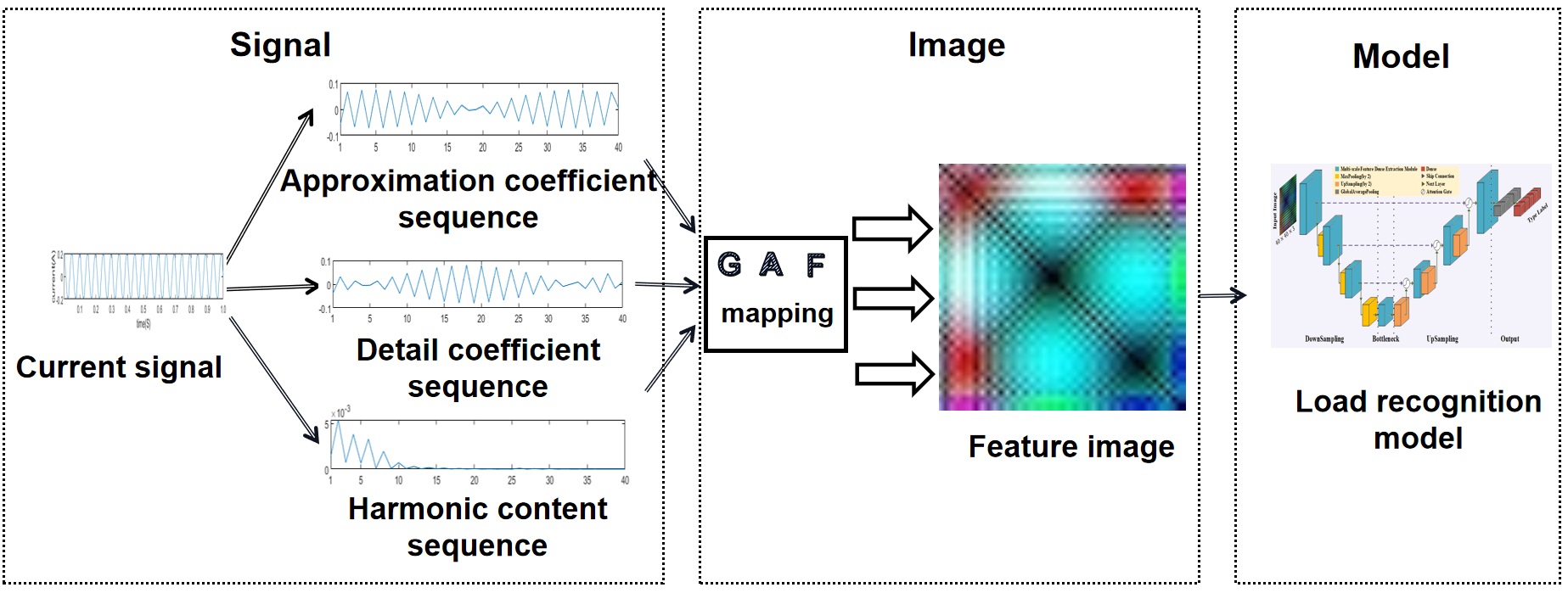}
	\caption{The overall framework of the proposed LRA.\quad\quad\quad\quad\quad}
	\label{figure_2}
\end{figure}

In the 
non-invasive design of Fig. 1 (b), Load Recognition Algorithm (LRA) plays an essential role and thus attracts attentions from researchers. Their contributions will be elaborated in Section \ref{rx}. Despite of these great efforts, there still exists a need to further improve the accuracy of LRA. Generally, load recognition may lead to inaccurate results when using inappropriate feature extraction, or false rejecting loads under masking effect – features of low power load are usually hard to be identified under high power loads. To address these issues, this paper presents a NILM method based on current feature visualization and a U-shape deep neural network for recognition. Our contributions are summarized as follows.

First, we 
propose a current feature visualization method based on signal transforms and Gramian Angular Field (GAF). By this operation, the feature differences between loads are highlighted to make ease of vision-based recognition.  

Second, we 
propose a U-shape deep neural network based on multi-scale feature extraction and attention mechanism. This design aims to further enhance the recognition accuracies and generalization abilities of our method, especially at low power conditions.     

Third, our 
NILM approach demonstrates its high efficiency in both public and our private datasets. To examine the generalization ability of proposed approach, we introduce a new dataset with 12 types of electric loads with powers from 24W to 1800W. Experimental results in this dataset as well as the public PLAID dataset validate our design.

The rest 
of paper is organized as follows. Section II reviews the related work on load monitoring. Sections III presents our contributions in current feature visualization and load recognition based on deep neural networks. Comprehensive experiments and analyses are presented in Section IV. Finally, Section V concludes this paper.

\begin{figure*}[htb]
	\centering
	\includegraphics[width=7in]{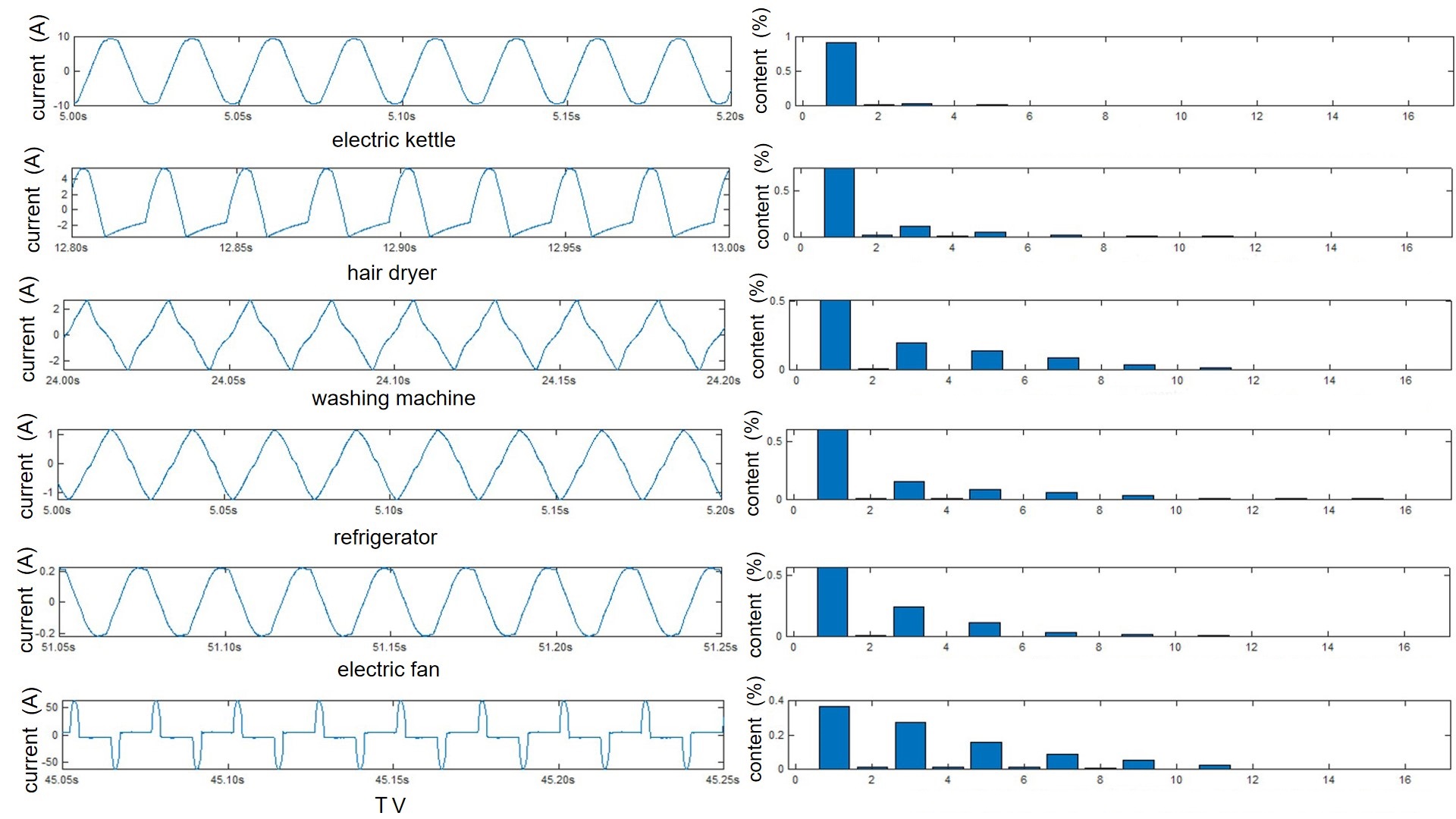}
	\caption{ Current signals and harmonic contents of typical loads.\quad\quad\quad\quad\quad\quad\quad\quad\quad\quad\quad\quad\quad\quad\quad\quad\quad\quad\quad\quad\quad\quad\quad\quad\quad\quad\quad}
	\label{figure_3}
\end{figure*}

\section{Related Works}\label{rx}

As pointed 
out in Section I, in a NILM method, only one terminal is deployed at the access point of family/cell. It sees the electrical loads in room within a black box. How to design an effective LRA model to recognize or interpret these loads is thus critical. 

Traditional LRA 
methodologies compared the feature of an unknown load with those of known loads in dictionary. They made judgments through a metric set consisting of matching degree \cite{9}, similarity degree \cite{10}, Hellinger distance \cite{11}, etc. The performance of LRA was also benefited from the development of machine learning, resulting in recognition methods with K-means clustering \cite{12} and fuzzy C-means \cite{13}.  However, these methods basically utilized single feature  without consideration on subtle differences between similar signals. Therefore, the problem of recognition confusion was not well addressed. 

Researchers considered 
to introduce more types of signal features to improve the accuracy of LRA. \cite{14} proposed a load recognition model with a feature combination of transient waveform and power change value during load switching. Kang {\it et al.} \cite{15} employed fast Fourier transform to extract the amplitude and phase of odd harmonics of the current, and then used them as key features for recognition. To improve the recognition accuracies of loads, \cite{16} constructed a hybrid feature set by the parameters of active power, reactive power and harmonic amplitude.

In the 
past decade, deep learning has demonstrated its strengths in AI-driven tasks, such as computer vision, natural language processing, human-computer interaction, and IoT. These successes also inspired the researchers to introduce deep neural networks in LRA. \cite{17} designed a sequence-to-sequence Long-Short-Term Memory (LSTM) network for load recognition. The authors in \cite{18} designed a capsule-network-based LRA, in which Convolutional Neural Network (CNN) extracted latent features from a set of non-overlapping energy measurement data segments. \cite{19} proposed a dual-stream neural network to extract features from current signals. \cite{20} proposed to extract features with Siamese neural networks and then used them in load recognition. These works have revealed the strong feature extraction abilities of neural networks with promising performance in load recognition.

To further
promote the LRA accuracy, researchers also attempted to visualize the features of voltage or current and then employed image-alike processing techniques in load recognition. Owing the advantage of recent booming of computer vision technologies, more accurate and robust LRA methods were developed. \cite{21} presented an image-classification-based LRA, where the image was obtained with voltage-current (V-I) trajectory. \cite{22} provided a CNN-based LRA with weighted pixel V-I trajectory map as features. Liu {\it et al.} employed color-coded V-I trajectory map as the input of their AlexNet-based load recognition model \cite{23}. In \cite{24}, the V-I trajectory and amplitudes of current and voltage were mapped as a color image, which provided richer feature information to CNN-based load recognition. Wenninger {\it et al.} \cite{25} mapped a cycle of V-I trajectories as a threshold-free recursive graphs, and subsequently designed a Spatial Pyramid Pooling (SPP) convolutional neural network for load recognition.

\begin{figure*}[]
	\centering
	\includegraphics[width=6in]{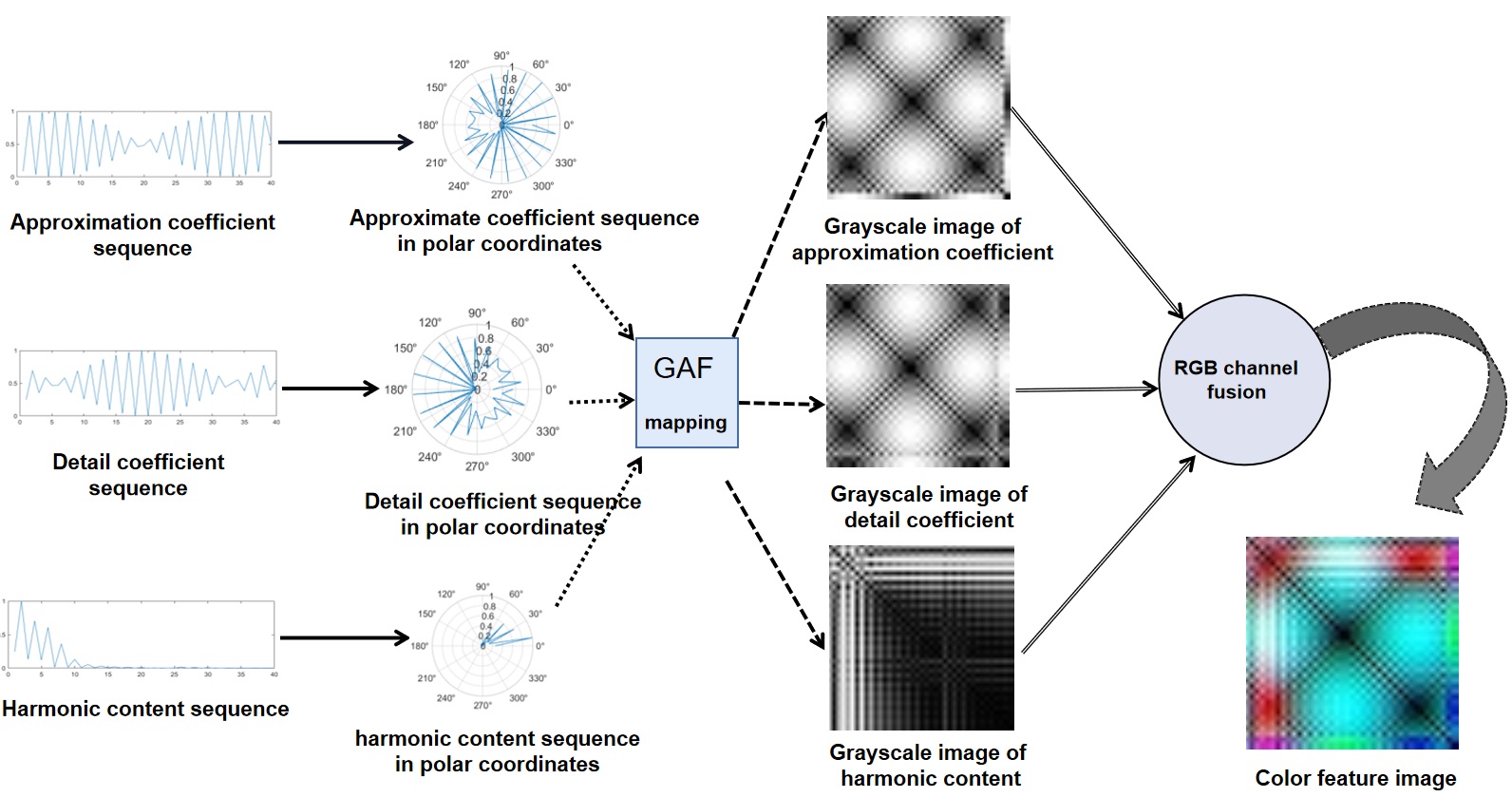}
	\caption{Flowchart of our proposed current feature visualization method.\quad\quad\quad\quad\quad\quad\quad\quad\quad\quad\quad\quad\quad\quad\quad\quad\quad\quad\quad\quad\quad\quad\quad}
	\label{figure_4}
\end{figure*}
Our proposed 
method is also an image-based recognition approach that maps current features and employs image-alike recognition. Compared with its peers, we design a more effective feature mapping framework and a more reliable deep neural network for load recognition. Experimental results validate its design.

\section{Proposed Method}
\subsection{Motivation and Our Framework} 
By summarizing 
the above analysis in Section II, we expect to design an effective LRA method for NILM using more detailed, visualized features and a finely-optimized deep neural network. Among them, more detailed and visualized features benefit load recognition, especially for similar loads and low-power loads masked by high-power loads. The finely-optimized deep neural network helps to improve the accuracy of LRA.

Inspired by this, we propose a NILM method based on current feature visualization and deep neural network, as shown in Fig. 2.  First of all, we employ wavelet and Discrete Fourier Transform (DFT) transform to decompose the current signal into three feature sequence: (i) approximation coefficient sequence with envelope feature of the current signal; (ii) detail coefficient sequence with texture feature of the current signal; (iii) harmonic ratio sequence with harmonic feature of the current signal.  After that, we utilize the GAF method to convert the above three sequences into gray-scale images, and further set them as R, G, B channels of a color image. By this operation, the differences between loads are highlighted to make ease of load recognition. Finally, we propose a load recognition model based on multi-scale features and visual attention mechanism. These steps will be elaborated as follows.

%\begin{figure*}[htb]
%	\centering
%	\includegraphics[height=3in,width=6in]{figure_2.jpg}
%	\caption{Framework of the proposed LRA.}
%	\label{figure_2}
%\end{figure*}

\subsection{Proposed Current Feature Visualization} 

In a 
NILM system, the terminal at commercial power input is able to capture the voltage and current signals. Between them, the voltage keeps almost intact while the current fluctuates with the electric usage on loads. Therefore, we select the current signal as the input of our load recognition model. To visualize the current signal as a two-dimensional image, we first extract its features and then convert them into gray-scale images, which are further set as channels of the visualized color image.

Denote the 
current signal by $c(n)$ with a length $N$. It can be expanded into wavelet series as \cite{26}:

\begin{equation}
\begin{aligned}
{c}{(n)} = \frac{1}{\sqrt{{N}}}{\sum_{k}}{ W}_\varphi(j_0,k)\cdot{\varphi_{{j_0},k}(n)}+\\
\frac{1}{\sqrt{{N}}}\sum_{j = j_0}^{\infty}{\sum_{k}}{ W}_\psi(j,k)\cdot{\psi_{{j},k}(n)},
\end{aligned}
\end{equation} 
where $j$ represents wavelet decomposition scale that determines the length of wavelet coefficient, $k$.  $\varphi_{j_0, k}(n)$ and $\psi_{j_0, k}(n)$ represent the scaling and wavelet functions, respectively. $W_\varphi(j_0, k)$ and $W_\psi(j, k)$ are approximation coefficient and detail coefficient, respectively. For ease of presentation, we denote them as $S_{\rm{a}}(n)$ and $S_{\rm{d}}(n)$, $n=0,1,2,..., N-1$. They are calculated as \cite{27}:

\begin{equation}
	\begin{aligned}
	{\rm S_a}{(n)} = \frac{1}{\sqrt{{N}}}{\sum_{n}}{ c}(n){\rm \varphi}_{{j_0},k}(n),
	\end{aligned}
\label{eq_2}
\end{equation} 

\begin{equation}
	\begin{aligned}
		{\rm S_d}{(n)} = \frac{1}{\sqrt{{N}}}{\sum_{n}}{ c}(n){\rm \psi}_{{j},k}(n),j>j_0.
	\end{aligned}
\label{eq_3}
\end{equation}

According to \cite{28}, $S_{\rm{a}}(n) $ and $S_{\rm{d}}(n) $ represent  the envelope feature and texture feature of the current signal, respectively. This paper uses them as two features that will be visualized and recognized. 

 Another feature used in this paper is harmonic content which refers to the percentage of the $k$-th order harmonic component in the total harmonic components. As the $k$-th order harmonic component of the current signal can be obtained by DFT \cite{29}:

\begin{equation}
	{r(k)}={\frac{1}{N}}{ \sum_{n=1}^{ N}}{ c}(n) e^{-\mathrm{\rm j} \frac{2{\rm \pi} k}{N} n} .
\end{equation}
the harmonic content (denoted by $S_{\rm{h}}(k)$ )  is formulated as:
\begin{equation}
	{S_{\rm{h}}}{(k)}=\frac{{ r}(k)}{\sum\limits_{i=1}^{N} { r}(i)} .
\end{equation}		
The harmonic contents of loads vary subject to electrical components and circuit systems. To verify the above assumption, we explored the harmonic contents of different types of loads: resistive, pump-driven, motor-driven and switching-powered. The results are shown in Fig. 3 where the left part shows the current waves while the right part shows their corresponding harmonic contents. From this figure, the resistive loads ({\it e.g.} electric kettle, hair dryer) heat with resistors and barely have harmonic components. The pump-driven loads ({\it e.g.} washing machine, refrigerator) works mainly with fundamental wave, but with more 3rd, 5th and 7th harmonic components. The motor-driven loads ({\it e.g.} electric fan) are similar to pump-driven loads. They have 3rd, 5th, 7th and 9th harmonic components that are lower than fundamental wave. The switching-powered loads ({\it e.g.} TV, computer) adjust output voltages with high-frequency switches. They generate rich high-order harmonic components, such as 3rd, 5th, 7th, 9th, 11th and 13th harmonic components, which are comparable with fundamental wave. Obviously, the harmonic content, {\it i.e.} $S_{\rm{h}}(k)$, is an effective feature to identify different types of loads. 

\begin{figure*}[htb]
	\centering
	\includegraphics[width=6in]{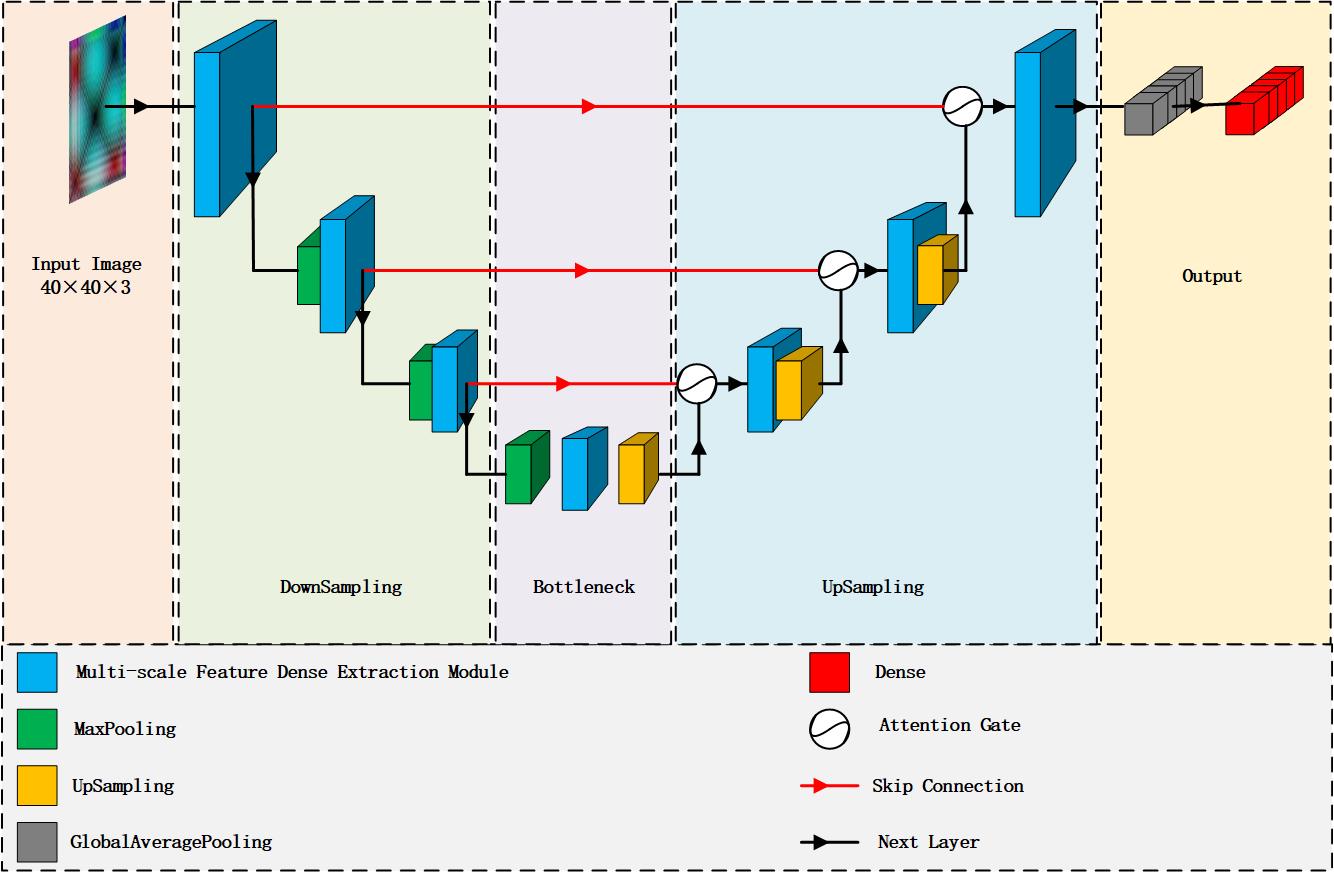}
	\caption{Proposed deep-learing-based load recognition network.\quad\quad\quad\quad\quad\quad\quad\quad\quad\quad\quad\quad\quad\quad\quad\quad\quad\quad\quad\quad\quad\quad\quad\quad\quad\quad\quad}
	\label{figure_5}
\end{figure*}
We then 
map all extracted features of $S_{\rm{a}}(n) $, $S_{\rm{d}}(n) $ and $S_{\rm{h}}(k) $ into grayscale images with GAF method. Take $S_{\rm{a}}(n) $ for example.  Firstly,  $S_{\rm{a}}(n) $, $n=0,1,2,..., N-1$ are transferred from Cartesian coordinate system to polar coordinate system:
\begin{equation}
	\left\{\begin{array}{l}
		{\rm \varphi}(n)=\operatorname{arctg}\left[{ S_{\rm{a}}}(n)\right] \\
		{ r}(n)=\sqrt{{ S_{\rm{a}}}^2(n)+n^2}
	\end{array} \right..
\end{equation}

Then, an $n{\times}n$ Gramian matrix $\boldsymbol {G}_{\mathrm{a}}$ is obtained in which
\begin{equation}
 \boldsymbol G_{\mathrm{a}}(i,j)=\sin(\varphi(i-1)-\varphi(j-1)) .
\end{equation}

Similarly, we 
also obtain the Gramian matrixes of ${ S_{\rm{d}}}(n) $ and ${ S_{\rm{h}}}(k) $, respectively denoted by $\boldsymbol {G}_{\rm{d}}$ and $\boldsymbol {G}_{\rm{h}}$. We use the following equations to map these Gramian matrixes to the R, G and B channels of a color image:

\begin{equation}
	\left\{\begin{array}{l}
		\boldsymbol {R}=\left|\boldsymbol {G}_{\rm{a}}\right| \times 255 \\
		\boldsymbol {G}=\left|\boldsymbol {G}_{\rm{d}}\right| \times 255 \\
		\boldsymbol {B}=\left|\boldsymbol {G}_{\rm{h}}\right| \times 255 
	\end{array}\right.
\end{equation}
\begin{equation}
	{\boldsymbol {I}_{\rm F}}=[\boldsymbol {R} \quad \boldsymbol {G} \quad \boldsymbol {B}] .
\end{equation}
The flowchart 
of our proposed current feature visualization can be summarized in Fig.~\ref{figure_4}.  The fused color image, ${\boldsymbol {I}_{\rm F}}$, with a resolution of 40$\times$40, has its unique texture information and chroma components. It includes low-frequency envelope features, high-frequency texture details and all harmonic ratios. Compared with the original one-dimensional current signal, it visualized and highlighted all hidden features whilst keeping timestamp of the original signal. As a result, we are allowed to use CNN methods to recognize all loads with a high accuracy.

\subsection{Proposed Deep Load Recognition Network} 
To recognize all loads from ${\boldsymbol {I}_{\rm F}}$, a critical issue is how to effectively extract both global and local features and avoid the influence of noises. To this end, we propose a deep load recognition network based on multi-scale features and attention mechanism. As shown in Fig.~\ref{figure_5}, the proposed network is a U-shape model with four stages: downsampling, bottleneck, upsampling and output.

\begin{figure*}[!t]
	\centering
	\includegraphics[width=7in]{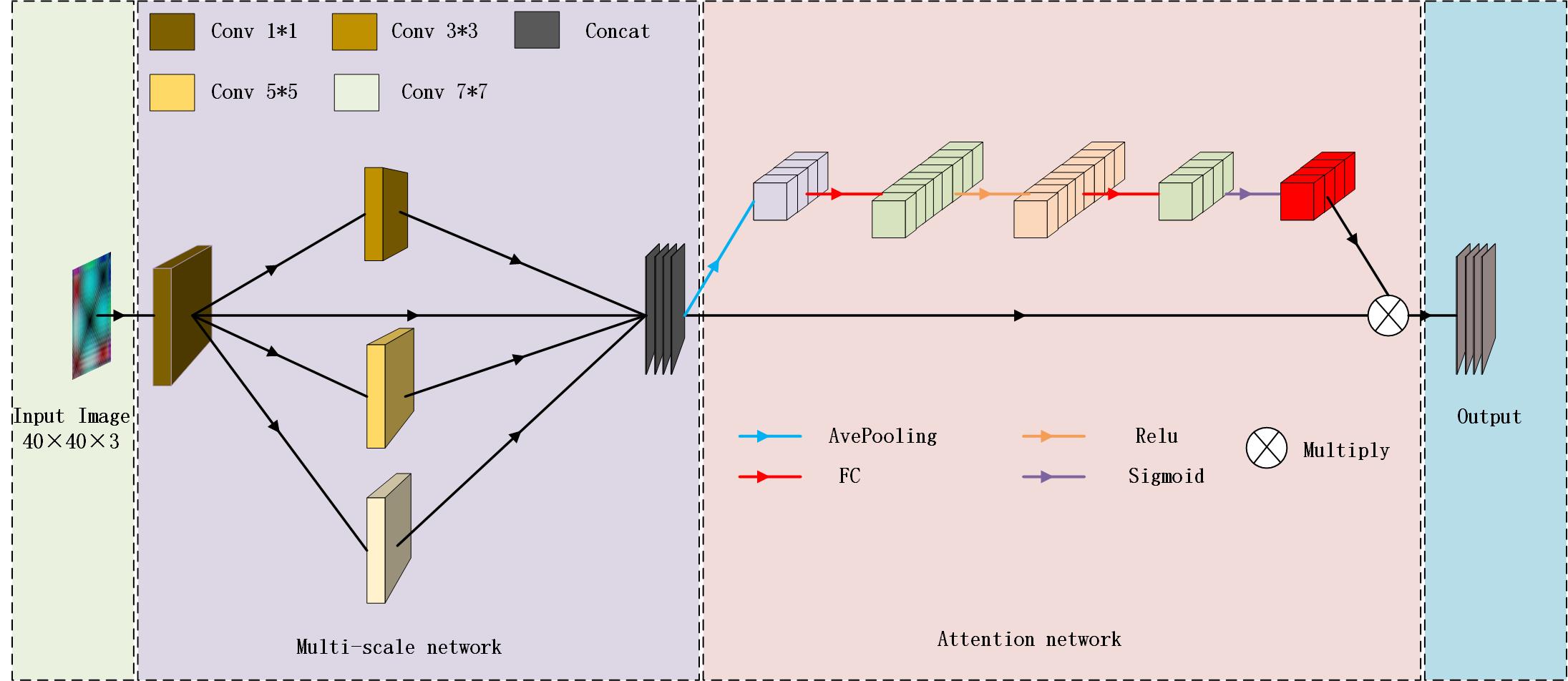}
	\caption{Proposed Network of MFDEM.\quad\quad\quad\quad\quad\quad\quad\quad\quad\quad\quad\quad\quad\quad\quad\quad\quad\quad\quad\quad\quad\quad\quad\quad\quad\quad\quad\quad\quad\quad\quad\quad\quad\quad\quad\quad\quad\quad\quad\quad\quad\quad\quad\quad\quad}
	\label{figure_6}
\end{figure*}
In downsamping 
stage, the inputted ${\boldsymbol {I}_{\rm F}}$ is sequentially processed by three attention-based Multi-scale Feature Dense Extraction Modules (MFDEMs) with max pooling operations by 2. We design the MFDEM with attention mechanism and multi-scale feature extraction, with its network structure shown in Fig.~\ref{figure_6}.  First,  it utilizes a 1$\times$1 convolution to extract coarse-grain features and further divides them into four feature sub-sets with the same space size. Second, it employs four different convolution kernels to extract features from different sub-sets: 

\begin{equation}
	b_i=\left\{\begin{array}{c}
		a_i \quad~~~~~~~~~~~~~~~ i=1 \hfill \\
		m_i\left(b_{i-1}+a_i\right) \quad 2 \leq i \leq 4
	\end{array}\right.
\end{equation}
where $m_2()$, $m_3()$ and $m_4()$ are 3$\times$3, 5$\times$5 and 7$\times$7 convolutions, respectively. Third, it concatenates all inputs $b_i, i\in [1,4]$ and uses a 1$\times$1 convolution to fuse all features from different scales. Fourth, it employs attention mechanism to adaptively allocate weights to multi-scale features. By using three MFDEMs, the proposed network deeply exploits the image features at different scales that benefit the feature extraction ability of our model.

The attention 
network uses an average pooling to obtain a feature map with global reception field to represent the importance of all channels. Then, it adds two fully connected layers and a ReLU function to introduce non-linearity in importance maps. Finally, it uses the Sigmoid function to generate new weights of channels and multiples them with the original features to obtain the attention-based features. Obviously, the weighting operation of feature maps highlights important features and depresses the other features, thus it degrades the impacts of non-relevant features in load recognition. 

In bottleneck 
stage, the downsampled features are processed by a max pooling, an MFDEM module and an Upsamling operation. Note that our network is a U-shape model that the output of each MFDEM is also fed to the upsampling stage with skip connection. This is to retain features of all scales, include both low-level and high-level features, which may benefit the load recognition to the extent possible. The bottleneck stage is set at the bottom of this U-shape model. 

\begin{figure}[!t]
	\centering
	\includegraphics[width=3.5in]{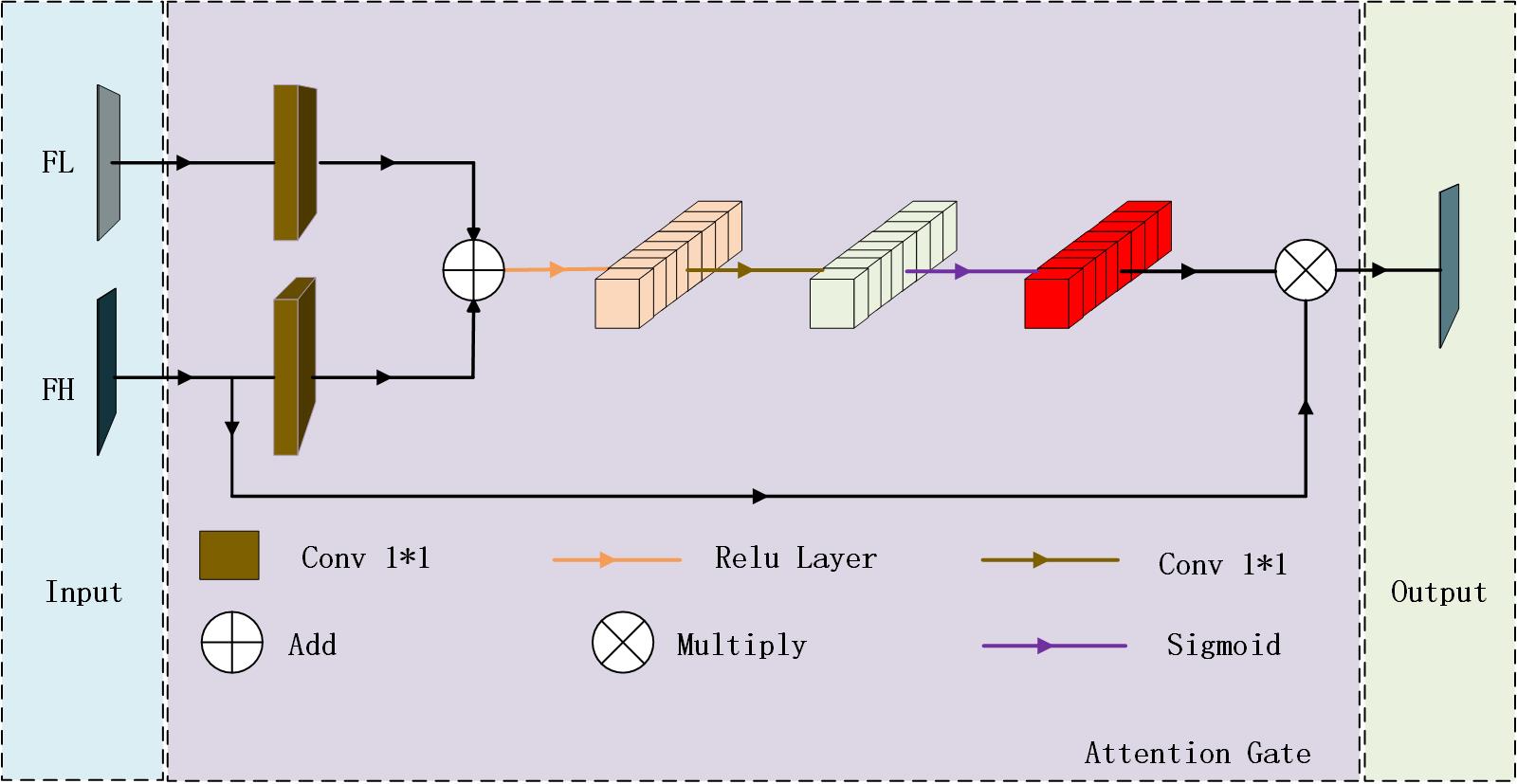}
	\caption{Network of AG.\quad\quad\quad\quad\quad\quad\quad\quad\quad\quad\quad\quad\quad\quad\quad\quad}
	\label{figure_7}
\end{figure}

In upsampling 
stage, the features are processed as an inverse path of downsampling stage that consists of MFDEMs and upsampling operations by 2. As discussed in last paragraph, skip connections are introduced to feed features from downsampling stage. However, a direct addition of these features would be ineffective in recognition. Here we introduce an Attention Gate (AG) before each addition operation to selectively enhance low-level features. Fig.~\ref{figure_7} shows the network structure of GA, where the low-level feature from skip connection (denoted by $F_L$) is set as the gate control signal, and the high-level feature (denoted by $F_H$) is used as the input signal. Both two signals are separately convoluted by 1$\times$1 and added into a new signal. Then, the new signal is activated by ReLU function, convoluted by 1$\times$1 and filtered by a Sigmoid function to obtain a weight matrix. Finally, the feature $F_H$ is multiplied with the above weight matrix to obtain a new feature $F_N$ that embodies both low-level and high-level features for load recognition. 

The last 
stage outputs load recognition results based on the fused multi-scale features. It utilizes a global average pooling, a fully connected layer and a Softmax function to classify all types of loads. The average max pooling operation can further increase the perception field for recognition and also output a smaller parameter size for the following calculation. The fully connected layer with Softmax is commonly used in classification and recognition tasks due to its effectiveness in non-linear regression.

\subsection{Smart Energy Management with Our Method}

In summary, the steps of proposed load recognition method are as follows.

Step 1. Obtain the current signal $c(n)$.

Step 2. Calculate the three feature sequences, approximation coefficient ${ S_{\rm{a}}}(n) $, detail coefficient ${ S_{\rm{d}}}(n) $ and harmonic content ${ S_{\rm{h}}}(k) $, with Eqs. (2), (3) and (5).

Step 3. Use the GAF method, as shown in Eqs. (6)-(9)  to convert these feature sequences into a color image ${\boldsymbol {I}_{\rm F}}$. 

Step 4. Identify the electric load with the image ${\boldsymbol {I}_{\rm F}}$ and the proposed deep network in Fig. 5.

\begin{figure*}[htb]
	\centering
	\includegraphics[width=6.5in]{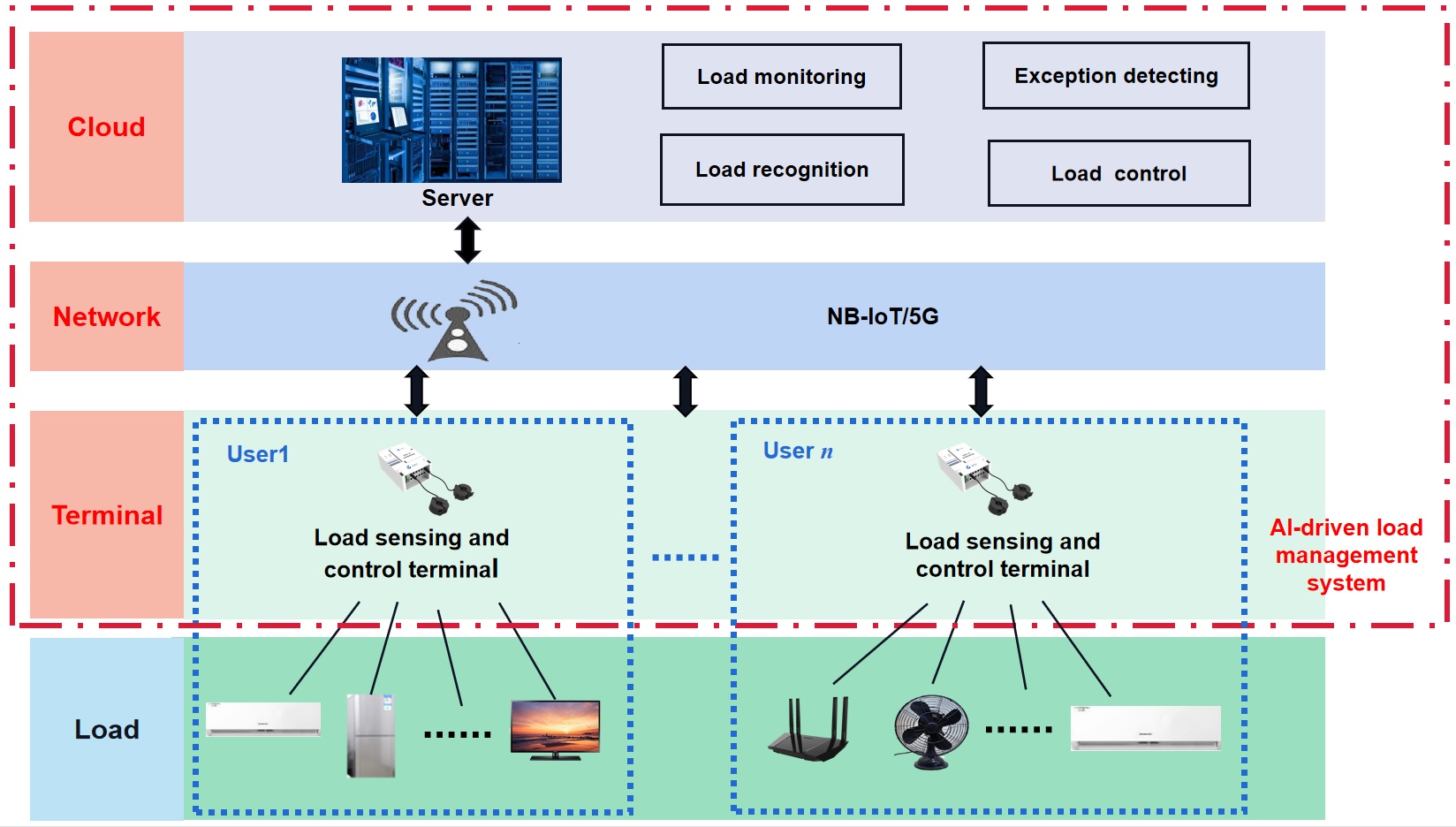}
	\caption{Framework of the proposed AI-driven load management system.\quad\quad\quad\quad\quad\quad\quad\quad\quad\quad\quad\quad\quad\quad\quad\quad\quad\quad\quad\quad\quad\quad\quad\quad\quad\quad\quad }
	\label{figure_8}
\end{figure*}

A practical AI-driven load management system can be implemented with our proposed load recognition method. As shown in Fig.~\ref{figure_8}, it is designed as a joint Terminal-Network-Could infrastructure for smart city. At the terminal-end, a load sensing and control terminal is in charge of managing all loads in a family or cell.  On one hand, it collects load information ({\it e.g.} the current signal) and sends them to cloud server for further analysis; on the other hand, it receives and executes all commands from cloud server to control all loads for energy saving. The network is utilized to transmit all data and commands. Generally, it can be designed based on Narrow Band IoT (NB-IoT) for its low-power design, or 5G for its high transmission speed. At the cloud-end, a cloud-based service  platform is responsible for load recognition, monitoring and control.  It checks and analyzes the electricity consumption of all loads and provides suggestions on smart control of switches.Through its control, we are able to avoid unnecessary ({\it e.g.} lighting under daylight) or dangerous ({\it e.g.} abnormal use in factory) usage of electricity loads. Through this design, we expect to support efficient electricity management over large-scale IoT in smart city.

\section{Experiments and Simulations}

To examine 
the performance of our method, we compare it with the state-of-the-art methods on popular datasets. We also present the ablation studies to validate the effectiveness of our network design.

\subsection{Datasets}

We compare 
all methods in a publicly available dataset, Plug Load Appliance Identification Dataset (PLAID) \cite{gao2014plaid}, with electric usage data of 11 types of loads. To increase the data samples for training, we also design a load sensing terminal and capture electric usage of 12 types of popular loads. The powers of these loads range from 24W to 1800W, which are representative to test the LRAs under similar electric loads or masking effect of loads.

\subsection{The Performance of Our Method}

\begin{figure}
	\centering
	\subfigure[Results on our dataset.]{
		\includegraphics[width=3.4in]{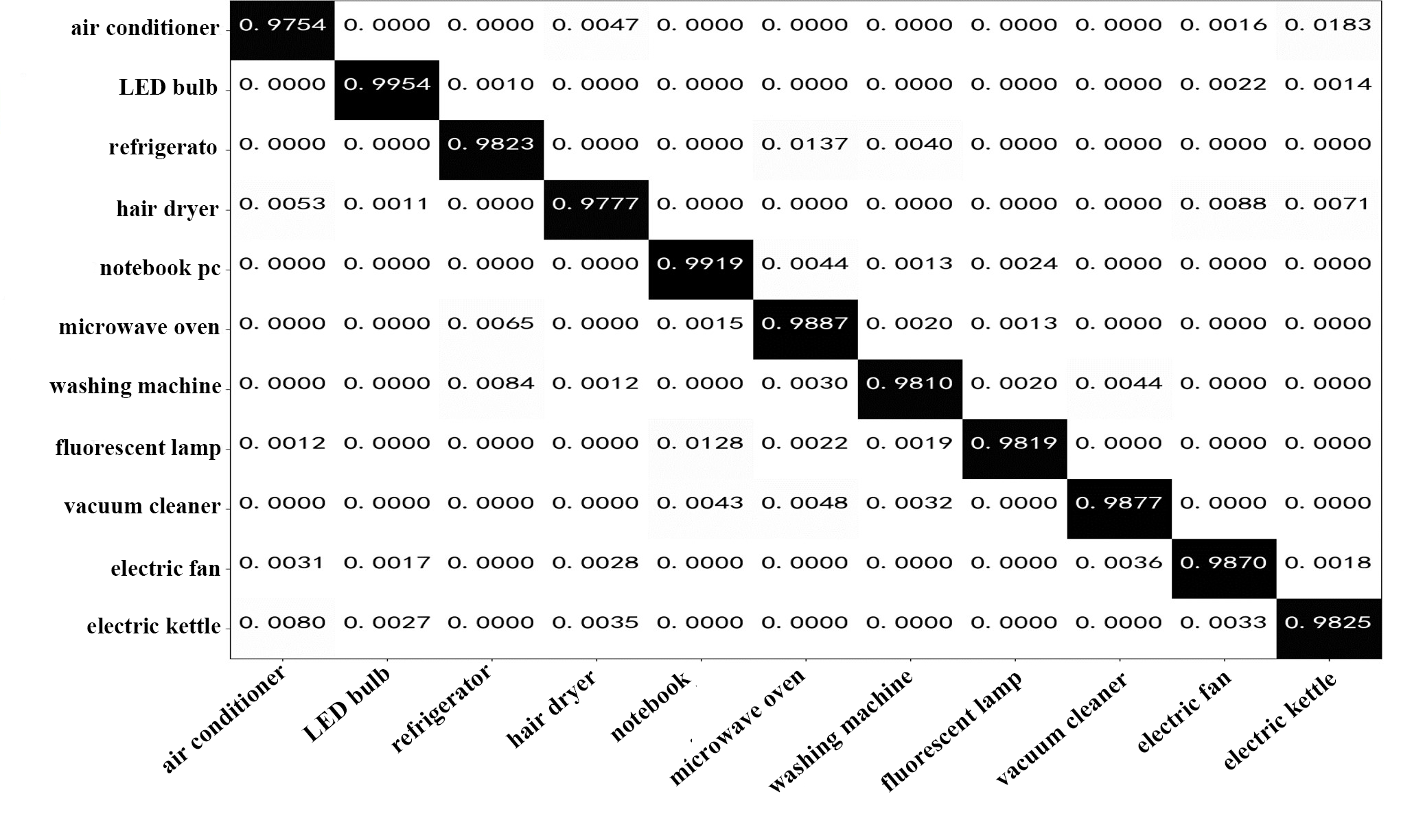}
		
	}
	
	\subfigure[Results on the PLAID dataset.]{
		\includegraphics[width=3.4in]{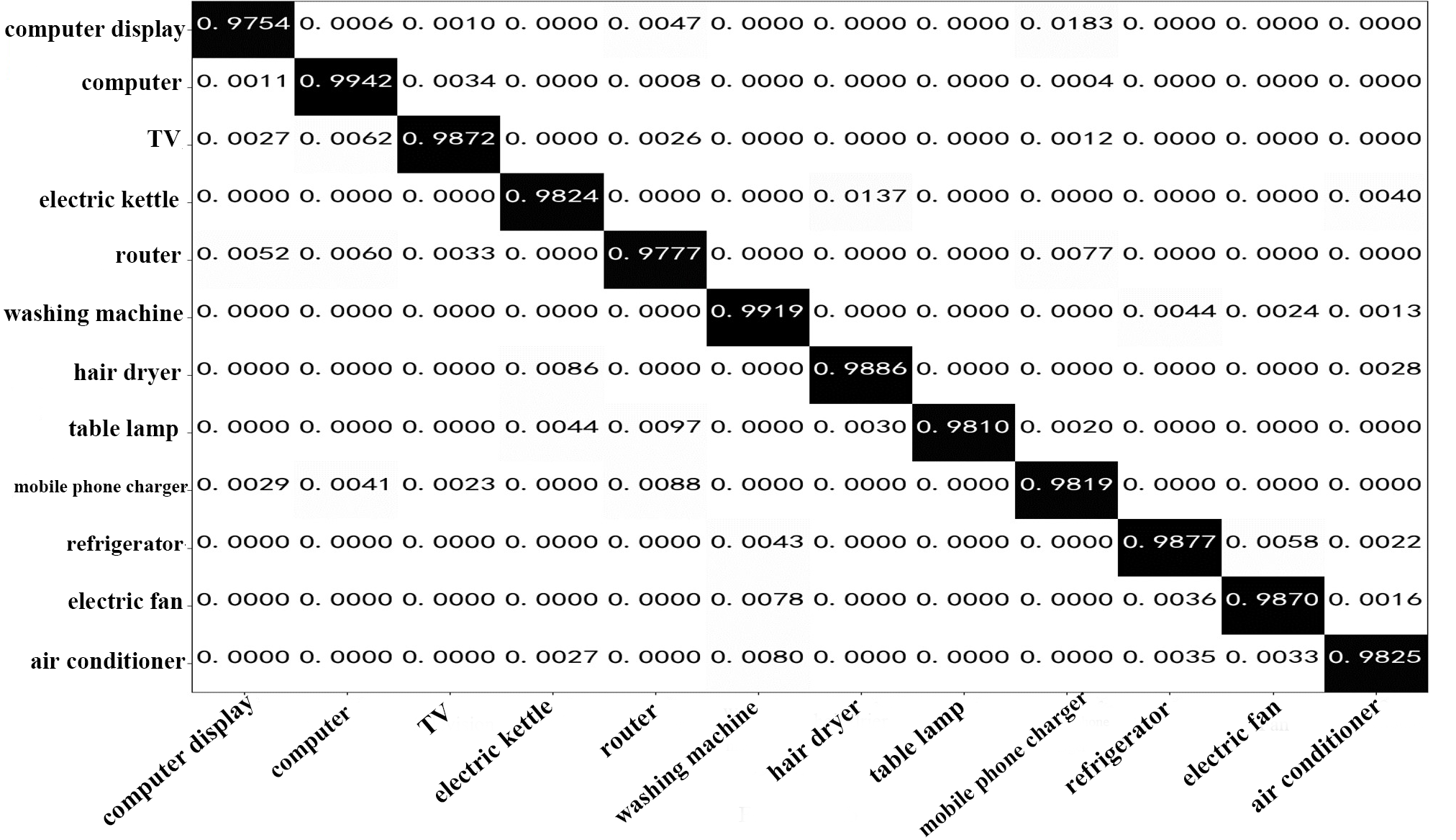}
		
	}
	
	\label{fig_9}
	\caption{Confusion matrixes of our method.\quad\quad\quad\quad\quad\quad\quad\quad\quad}
\end{figure}
The tested 
results of our method are summarized in Table \uppercase\expandafter{\romannumeral1}, including results on both PLAID and our datasets. For each dataset, we split it into a 70:30 ratio for training and testing. From the table, our method achieves a promisingly high accuracy of 98.26\% on our dataset with an F1 score of 0.9819. It also achieves an accuracy of 97.71\% and an F1 score of 0.9743 on the PLAID dataset. These results have fully demonstrated the efficiency of our load recognition model.
\def\tablename{TABLE}
\begin{table}[h]
	\begin{center}
	\caption{Performance of our method}
	\label{tab:tracking}
\begin{tabular}{ccccc}% 
	\toprule %[2pt] 
Database &Accuracy&	Precision&	Recall	&F1 \\ %
	\midrule %[2pt]  
Ours&	0.9826&	0.9818&	0.9821	&0.9819\\\\
PLAID	&0.9771&	0.9779&	0.9707&	0.9743\\
	\bottomrule %[2pt]     
	
\end{tabular}
\end{center}
\end{table}

To analyze the recognition accuracies of different types of loads, we present the confusion matrixes on testing sets in Fig. 9. From the figure, all resistive loads, {\it e.g.} electric kettle (1800W) and hair dryer (1200W), can be well recognized with a high accuracy of 98.20$\%$. This fact is mainly attributed to their distinct features of high powers and low harmonic contents. As discussed in Section III.B, pump-driven loads have similar feature distributions to those of motor-driven loads. In these categories, our model successively identifies washing machine, electric fan, air conditioner and vacuum cleaner with a low probability of incorrect recognition or confusion. switching-powered loads have high harmonic contents, but also with low powers that might be covered by high-power loads. From the figure, our method is well capable of addressing this issue. Its recognition accuracies of notebook, TV and router exceed 96.88\% even when high-power loads ({\it e.g.}, electric kettle, air conditioner) are working at full powers.

\subsection{Comparison with Popular Methods}
To examine 
the superiority of our proposed method, we compare it with references [22-24] under the same conditions. The evaluation results on PLAID dataset are shown in Table \uppercase\expandafter{\romannumeral2}. From the table, our method surpasses all compared algorithms on PLAID with an average accuracy of 97.71\%. This fact validates the effectiveness of our method in electric load recognition. It is thus capable of identifying all types of loads after a fine training in a large-scale electricity management system for smart city. 
\begin{table}[h]
\begin{center}
\caption{Comparison results on the PLAID dataset}
\label{tab:tracking}
\begin{tabular}{ccc}
\toprule
	%[2pt]    
Method &  Network & Accuracy(\%) \\ %
\midrule
	  %[2pt]  
De’s [22]    &       CNN        &   91.74    \\\\
	      Liu’s [23]        &     AlexNet      &   95.40    \\\\
	      Ding’s [24]       &       CNN        &   96.63    \\\\
	         Ours          & U-shape &   97.71    \\ \bottomrule
	        %[2pt]         &
\end{tabular}
\end{center}
\end{table}

In addition, both the two stages of our model, current feature visualization and deep load recognition, contribute to the final performance. With our current feature visualization and Ding’s CNN model, the hybrid method also achieves an accuracy of 96.91\% that outperforms Ding’s method. By using both current feature visualization and deep load recognition, our method outperforms Ding’s by 1.08\%. This fact also validates our design.

\subsection{Ablation Study}
We also 
perform ablation experiments to examine the design of our deep load recognition network. Experiments are run on both PLAID and our dataset.

\textit{The effectiveness of multi-scale feature extraction}: Our deep network utilized multi-scale convolution kernels {1$\times$1, 3$\times$3, 5$\times$5, 7$\times$7}, as shown in Fig.~\ref{figure_6}. To validate its effectiveness, we compare it with identical kernel  settings, {\it e.g.} 1$\times$1, 3$\times$3, 5$\times$5 or 7$\times$7, with results summarized in Table \uppercase\expandafter{\romannumeral3}. All settings are retrained for fair comparison. From the table, our multi-scale feature extraction design is superior to all other settings. Therefore, our deep model can well extract all critical information from the visualized current and thus is more suitable for load recognition.
\begin{table}[h]
\begin{center}
	\caption{Performance evaluation with different kernels}
\begin{tabular}{ccccc}% 

	\toprule %[2pt]
Kernel &	Accuracy&	Precision &	Recall &	F1 \\
\midrule %[2pt]  
1$\times$1	&0.85100&	0.85611&	0.85076	&0.85342\\\\
3$\times$3	&0.92458&	0.92573	&0.92184	&0.92378\\\\
5$\times$5	&0.95333&	0.95576&	0.94941	&0.95257\\\\
7$\times$7	&0.95497&	0.95720	&0.95393	&0.95566\\\\
Ours&	0.98465	&0.98681&	0.98114	&0.98396\\
	\bottomrule %[2pt]     

\end{tabular}

\end{center}

\end{table}

\emph{The effectiveness of Attention Gate}: The AGs are utilized to add multi-scale features after skip connections. From Table \uppercase\expandafter{\romannumeral4}, our method achieves inferior performance without these AGs. Therefore, both multi-scale feature extraction and attention mechanism contribute to the final recognition performance. 
\begin{table}[h]
	\begin{center}
		\caption{Performance evaluation w/ and w/o AGs}
		%\resizebox{90mm}{8mm}
		{
		\begin{tabular}{cccccc}% 
			
			\toprule %[2pt]
			 &	Parameters	&Accuracy &	Precision &	Recall &	F1\\
			\midrule %[2pt]  
			w/ AG&	1264014&	0.98465	&0.98681&	0.98114&	0.98396\\\\
			w/o AG	&1247515&	0.97497	&0.97481&	0.97235&	0.97357\\
			\bottomrule %[2pt]     
			
		\end{tabular}}
	\end{center}
\label{label_4}
\end{table}

\subsection{Discussion on Practical Use}
As shown in Fig. 8, our method can be deployed at the cloud-end of smart city. The system embodies a joint Terminal-Network-Cloud infrastructure where a load monitoring and management system that surveils power on/off and electric usage of loads for each family/cell, and also controls the electric usage if necessary. At the terminal-end, the system captures all required information of electric usage; while at the cloud-end, the system identifies all working loads, analyzes the regional or local electric consumption and sends controlling commands if necessary. Note the could-server runs our NILM method only if necessary. Thus, the computational cost is not a critical issue. Actually, our experiments also show that we are able to run this NILM approach on a laptop, which is reasonable because our deep model processes a 40$\times$40 color image only. As a result, the proposed NILM approach can be deployed in the smart electricity management system of smart city.

\section{Conclusion}
This paper proposes a NILM method based on current feature visualization and deep recognition network. It first converts current signals into color images with a combination of signal transform and GAF method. Then, it utilizes multi-scale feature extraction and attention mechanism to design a U-shape recognition network. Experimental results demonstrate its efficiency in both public and our own datasets, as well as its performance superiority compared with its peers. The proposed method can be embedded into the load monitoring and management system to control the electric usage and save energy in smart city. 
%\clearpage
\bibliographystyle{IEEEtran}
\bibliography{reference}

\flushend

\end{document}